# An Architectural Model for a Grid based Workflow Management Platform in Scientific Applications

**Alexandru Costan\*, Florin Pop\*, Corina Stratan\*,
Ciprian Dobre\*, Catalin Leordeanu\*, Valentin Cristea\***

*\*Faculty of Automatic Control and Computers, University Politehnica of Bucharest
( e-mail: alexandru.costan , florin.pop, corina.stratan, ciprian.dobre, catalin.leordeanu, valentin.cristea@ cs.pub.ro).*

**Abstract:** With recent increasing computational and data requirements of scientific applications, the use of large clustered systems as well as distributed resources is inevitable. Although executing large applications in these environments brings increased performance, the automation of the process becomes more and more challenging. While the use of complex workflow management systems has been a viable solution for this automation process in business oriented environments, the open source engines available for scientific applications lack some functionalities or are too difficult to use for non-specialists. In this work we propose an architectural model for a grid based workflow management platform providing features like an intuitive way to describe workflows, efficient data handling mechanisms and flexible fault tolerance support. Our integrated solution introduces a workflow engine component based on ActiveBPEL extended with additional functionalities and a scheduling component providing efficient mapping between tasks and available resources.

## 1. INTRODUCTION

Modern scientific and business applications requirements stress the need to compose Web services, which results in developing composite Web services. Simply put, composition is about making independent Web services interact with one another according to a specific (application oriented) logic. Workflow technology fosters a two-levels programming model: at the lower level, (business or scientific) activities are defined and implemented modularly (potentially by distributed systems), and in a flow-independent way; on the upper level the flow of the process is defined by composing these activities. Major benefits of this approach are flexibility, reusability, scalability and integrability. Workflow management systems (WFMSs) (Leymann et al., 1997) provide the foundation for defining, automating, and executing flexible business processes.

Different specification languages of composition exist including the Business Process Execution Language (BPEL) (Andrews et al., 2003) and the Web Service Choreography Interface (WSCI) (Arkin et al., 2002). BPEL is an open specification for describing and executing flexible WS-flows (workflows where the only allowed participants are Web services). This specification is strongly supported by the industry and academia, and is becoming the standard for defining WS-flows.

Although BPEL is the de facto standard for Web services composition, current engines that implement BPEL specifications suffer from major limitations, especially for scientific applications. Indeed, even though a number of open source workflow systems are available, many of them are too difficult to use for non-specialists (some of them lack a graphical interface), or are restricted to a specific type of applications or on a single middleware platform. Moreover, to use BPEL for defining WS-flows pays in terms of flexibility. According to the workflow technology, processes are not coded into the systems but defined as separated entities (two-level programming model). Additionally, Web services technology provides for reusability and interoperability. The resulting combination is aligned with the current requirements on enterprise and scientific distributed systems (i.e flexibility to adapt to ever-changing requirements). Clearly, BPEL provides many benefits in the direction towards the development and execution of flexible business processes. However, there is still room for improvement and all these problems have been impeding the adoption of workflow-based solutions in the scientific community.

This makes designers envisage alternatives to address all possible issues related to mapping workflows to scientific applications. Our purpose is to develop a workflow management platform for distributed systems, targeted to scientific applications, that will provide solutions for the following aspects:

• an intuitive way to describe workflows, based on ontologies specific to the application domains, allowing the user to work with abstract components

• flexible workflow structure, allowing the orchestration of services and also of plain executable programs as well as adapting (managing) running processes

- efficient mechanisms for data handling, as scientific applications usually produce significant amounts of data; the mechanisms will be based on the data replication services provided by the underlying middleware

- comprehensive fault tolerance support, with configurable policies; as semantics and side effects vary from one application to another, we believe that the users should be able to select from multiple fault tolerance approaches the one that is the most suitable for a particular workflow

We provide an architectural model for a scientific workflow platform and incorporate it into an existing open-source BPEL engine (ActiveBPEL). The resulting system, an enhanced BPEL engine, supports the execution of standard BPEL processes as well as processes defined with an enhanced semantic syntax.

The remainder of this paper is organized as follows: Section 2 introduces the workflow platform architecture; Section 3 and 4 detail its major components: the workflow engine and the scheduling system. Section 5 presents the related work in the field of workflow engines using Grids and in Section 6 we summarize the conclusions of this study and the future work directions.

## 2. THE WORKFLOW PLATFORM ARCHITECTURE

The workflow management platform will have three main components: a high-level module that will provide a user interface for defining abstract workflow, and that will manage domain-specific ontologies; a middle-level module that will have the role of a workflow engine; and a low-level module that will be in charge of scheduling the workflow activities and services onto the distributed system's physical resources, relying upon the available middleware.

Our focus in this work is on the middle-level module, the workflow engine (highlighted in green in Figure 1), and on the latter, scheduling component. We have started by studying the facilities offered by the most commonly used workflow engines for scientific applications, from the point of view of the requirements presented above. Although some workflow engines provide advanced features for abstract workflows, data management or fault tolerance, they lack functionality in what concerns the other aspects.

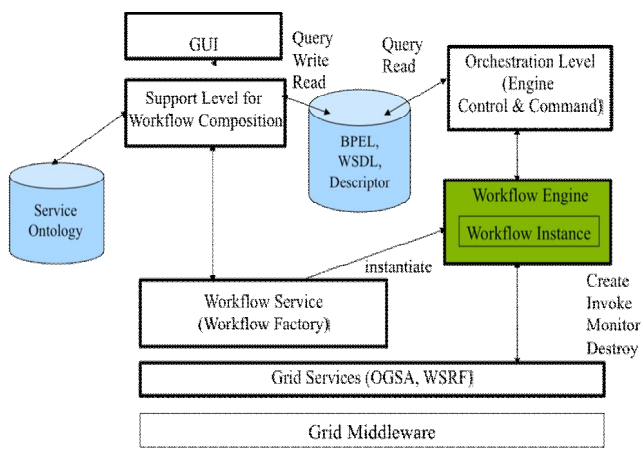

Fig. 1. Workflow Platform Architecture

As a consequence, we consider the approach of starting from an existing open source workflow engine and implementing additional functions that are required for the purposes of our project.

The engine we have studied is ActiveBPEL, one of the most widely used engines for WS-BPEL. We introduce here an architectural model of the modified ActiveBPEL engine, augmented with a new set of modules that will implement the additional functions.

## 3. WORKFLOW MANAGEMENT ENGINE

As we have shown in the previous sections, although several open source workflow engines are available for executing scientific applications in distributed environments, most of them lack important features concerning fault tolerance, abstract workflows, data handling and user interface. We note however that some of the existing engines are based on highly expressive languages and provide advanced process management, transaction handling, database persistence and other mechanisms. As a consequence, we chose the solution is of starting from an open source workflow engine and building additional modules to satisfy our requirements.

The workflow language we propose for the platform is WS-BPEL, which is a widely adopted standard in industry and, more recently, in academic environments. In what concerns the base workflow engine, we propose ActiveBPEL, which is the most frequently used open source BPEL engine and has been integrated in several research projects; some of the projects, like the one presented in (Subramanian et al., 2008), also augmented WS-BPEL with additional modules. We briefly describe as follows the ActiveBPEL architecture and the extensions we intend to implement for our project.

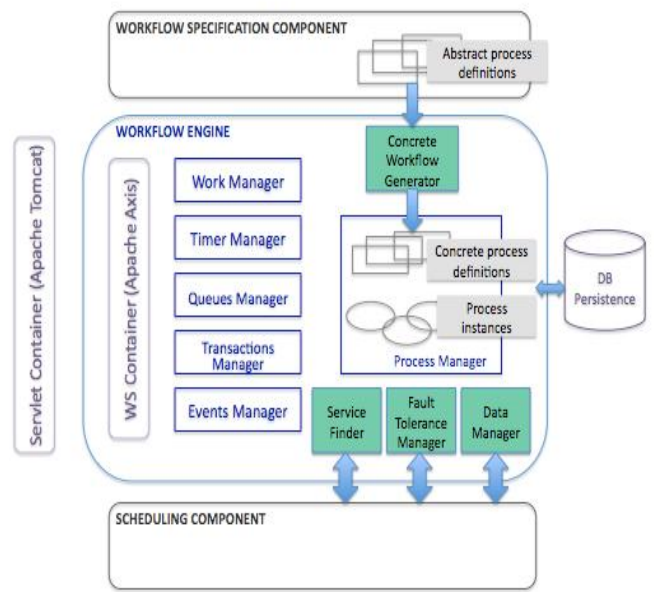

Fig. 2. ActiveBPEL-based workflow engine architecture. The new modules with which we propose to extend ActiveBPEL are depicted in green.

ActiveBPEL runs on top of the Apache Tomcat servlet container, and uses an embedded version of Apache Axis for message communications. Figure 2 presents the main components of ActiveBPEL (in blue) and our proposed extensions (in green). Among the services used in ActiveBPEL for handling processes, which are named Managers, the most important one is the Process Manager. The Process Manager oversees the instantiation and execution of processes and activities. When a process is deployed, the engine analyzes the BPEL sources and generates an internal representation of the process; then, when the user requires the execution of the process, a new instance is created by the Process Manager. The Process Manager is also responsible with instantiating activities and associating them with states (inactive, executing, finished, faulted etc.) during their life cycle. The Queue Manager handles incoming messages and events addressed to the process activities, by building a queue with the activities that are waiting for messages. The Work Manager schedules asynchronous operation, based on "work objects" which are a specialized alternative to threads. We also mention the Time Manager, which provides support for timed operations (like suspending or waiting), and the Transaction Manager, which implements methods for working with transactions.

We propose to introduce the following new components in the ActiveBPEL engine:

• Concrete Workflow Generator, which will transform abstract workflows into concrete workflows

• Service Finder, which will map service port types with sets of corresponding available services

• Fault Tolerance Manager, which will apply the policies specified by the user for handling faults

• Data Manager, which will implement efficient data handling mechanisms.

*3.1 Support for Abstract Workflow Specification*

The Concrete Workflow Generator will have as inputs the abstract workflows specified by the users with the aid of the Workflow Specification component; its role will be to perform the mapping between abstract functional components and web service port types or executable programs. As we focus on scientific applications we are concerned about the particular design of these workflows which typically requires the involvement of at least two domains: one from the scientific field of interest (e.g. high energy physics, molecular biology) and another from computer science - understanding the process of composing the workflow and encoding the derivation in a format that the engine can execute. Because these domain have distinct terminology to describe workflow elements, including requirements, clear specification and effective mapping are a challenge.

Our approach for abstract workflow specification uses ontologies, as they are used to describe knowledge about a domain such that its representation can be interpreted and reaseoned about by a computer. We use ontologies first as an explicit specification of abstract concepts and later to support the composition and matching of services.

While domain expert's workflow descriptions are more often abstract, our engine needs a concrete specification of an executable workflow. We therefore opted for the use of BPEL for Semantic Web Services (BPEL4SWS) (Nitzsche et al., 2007) as a means to increase productivity during the design of workflows in support of scientific applications. BPEL4SWS introduces the desired level of abstraction for modeling workflows that is consistent with the target domain. It is thereby used by our Concrete Workflow Generator component to automatically generate executable workflows, that is, workflow implementations.

In our proposed architecture, BPEL4SWS uses Semantic Web Service Frameworks to define a communication channel between two partner services instead of using the partner link which is based on WSDL 1.1. It enables describing activity implementations in a much more flexible manner based on ontological descriptions of service requesters and providers. The specification introduces an extension to BPEL to enable describing interaction using semantic Web service Frameworks instead of using WSDL 1.1. Semantic Web services (SWS) can be considered an integration layer on top of Web services; they use ontologies as data model and they have a rich conceptual model. There are efforts towards standardizing this conceptual model within the Reference Ontology for Semantic Service Oriented Architectures (RO4SSOA). In addition to the SWS based interaction, BPEL4SWS makes use of annotated data types to enhance data handling by means of ontological mediators and uses ontological reasoning to evaluate conditions. Our Concrete Workflow Generator component receives a BPEL4SWS specification as input and translates it into WS-BPEL, used by the ActiveBPEL engine to execute the workflow.

*3.2 Dynamic Web Services Composition*

The Service Finder will contact the scheduling component in order to discover web services (ports) that correspond to the port types specified in the workflow, using a find-and-bind approach similar with the one presented in (Miles et al., 2007). We aim at transparently adapting existing composite services to encapsulate autonomic behavior (Kephal et al., 2003). That is, making composite services adaptable to changes in their execution environment (e.g., failure in a partner Web service). Although this is a major concern in the field of composite services, it is often not addressed in the specification of composition languages.

Our Service Finder component maps the abstract nodes onto matching services iteratively during the processing of the workflow. Each time the workflow engine reaches a transition related to an abstract (non-executable) operation, it calls a special workflow refinement service. This service refines the workflow description by searching for matching service candidates, which fulfill the requirements defined by the profile of the abstract nodes. The decision of whether a service matches the requirements is done by rules that depend on several properties, such as functionality (e.g., service produces certain class of output data or side effect), performance (e.g., operation should complete within 1h), or reliability (e.g., only services which have been operational during the last 72h should be taken into account). If it is possible to find matching service candidates, the refinement service attaches a list of the corresponding interface descriptions URLs (e.g. wsdl URLs) to the abstract transition. Next, the binding consists of the selection of one service instance out of the list of available service candidates at runtime. In order to optimize this dynamic selection, the system uses input from the scheduling component of the platform, which takes into account the recorded as well as the current monitoring information about the services and the Grid infrastructure.

*3.3 Fault Tolerance*

Scientific processes are by nature long running processes. The execution of some kind of processes is expected to take days or weeks. Certainly, during this time, applications' conditions might change. However, at the moment, BPEL provides no special-purpose mechanism for supporting change on running processes or error support, and the only approach involves the tasks of stopping, un-deploying, re-designing and re-deploying the process. The problem is that to stop a running process in order to adapt it to a new situation is not cost-effective. In fact, the cost of stoping a running process is potentially high. Stopping a scientific process could affect the work that is being (and has been) done in several organizations. Either way, these processes cannot be simply stopped. In general, the development and execution of a set of roll-back activities to compensate the abortion of the process is required. This approach constraints the ability of a scientific application to adapt its processes to new situations. In this context, the need of introducing adaptability and fault tolerance mechanisms into the BPEL meta-model (in support of more adaptable processes) or directly into the workflow engines is recognized.

Indeed, the Fault Tolerance Manager's role is to attempt the recovery after an activity failure, by applying one of the available policies: re-try the activity, find an alternative service to invoke, save the partial results; activity replication is another approach that the user will be able to choose. A significant drawback of existing workflow systems is their poor support for exception handling. Our component aims at identifying the specific error conditions which occurred and taking consequent actions. Hence, the Fault Tolerance Manager distinguishes and reports the exceptions to which is confronted: failures of invoked applications, communication failures, lack of response from a user, missed deadlines, and unexpected behavior of applications. This is achieved by subscribing to the Queue Manager of the ActiveBPEL engine and inspecting all error related messages.

The usual failure-handling procedure in most systems is to stop process execution and report the failure to an administrator. However, as workflow applications become larger and more complex, manual failure resolution becomes less and less feasible because of the demand for human resources, with their high cost and slow answer time. Clearly, we need automatic exception handling, especially for scalable systems. Therefore, our approach automatically applies the hierarchy of policies defined by users. Thus, the Actions module of the Fault Manager component is able to take action when some configurable condition is met. This way, when a given threshold is reached, an alert e-mail can be sent, or a program can be run, or an instant message can be issued. Actions represent the first step towards the automation of the management decisions in scientific workflows.

*3.4 Data Management Functionalities*

For the efficient management of the workflow data, we propose to introduce the Data Manager component, which will contact the underlying middleware in order to find mappings between logical and physical file names, and will generate metadata that will allow making associations between files and the applications that produced or modified them. We extend ActiveBPEL with disk usage optimization techniques by implementing the algorithm presented in (Ramakrishnan et al., 2007). Hence, we minimize the disk space footprint of scientific workflows by removing data as soon as it is no longer needed and scheduling the workflow tasks by first taking into account the data requirements of the workflow and the data space availability at the resources.

## 4. THE DYNAMIC RESOURCE ALLOCATION (DyAG) COMPONENT

This component is concerned with workflow scheduling under resource allocation constraints. The resources in a workflow environment are agents such as machines, software, etc. that execute the task. Execution of a task has a cost and this may vary depending on the resources allocated in order to execute that task. Resource allocation constraints define restrictions on how to allocate resources, and scheduling under resource allocation constraints provide proper resource allocation to tasks. In this section, we provide an architecture to specify and to schedule workflows under these resource allocation constraints. The main components of this architectural model are the Dynamic Resource Allocator and the Fault Tolerance Manager which are discussed in the following.

*4.1 Dynamic Resource Allocation*

The scheduling in Grid systems is very complicated. The resource heterogeneity, the size and number of tasks, the variety of policies, and the high number of constraints are some of the main characteristics that contribute to this complexity. The necessity of scheduling in Grid is sustained by the increasing of number of users and applications. The design of scheduling algorithms for a heterogeneous computing system interconnected with an arbitrary communication network is one of the actual concerns in distributed system research. The resource allocation in dynamic way in Grid Environment is a complex problem. When talking about the design of a schedule for a given system and implicitly of a scheduling algorithm must consider all aspects of the system and applications that will run in it. In the case of dynamic scheduling, the basic idea is to perform resource allocation on the fly while other applications are in execution. This is useful in the case where jobs arrive in a real-time mode. Dynamic scheduling is usually applied when it is difficult to estimate the cost of applications, or jobs are coming on-line dynamically (in this case, it is also called on-line scheduling). A good example of these scenarios is the job queue management in some meta-computing systems like Condor and Legion. Dynamic task scheduling has two major components: system state estimation (other than cost estimation in static scheduling) and decision making.

It the services are considered, the process became more complicated. The resource managed is not able to invoke the service, but the endpoint to the services could be established. In this case, the policies are very important for services selection. The dynamic allocation has an important role for such component.

The Dynamic Resource Allocation (DyAG) module allows for the use of various policies and algorithms. This module is developed like a web service in gLite. The simplest of these policies and the first one to be tested is FCFS (First Come First Served). The resource allocation component is based on the solution presented in (Pop et al., 2008). To test the DyAG module two workflow applications were chosen: Gromacs (Lindahl et al., 2001) and Montage (Berriman et al., 2004). A large number of test scenarios will be developed based on the workflows generated by these two applications.

The component will run under Globus and gLite Grid middleware and will be responsible for the dynamic resource allocation in the Grid system, as well as for the failure management and workflow scheduling. The main components of this module and its interface to the Workflow Engine described in Section 2 are displayed in Fig. 3.

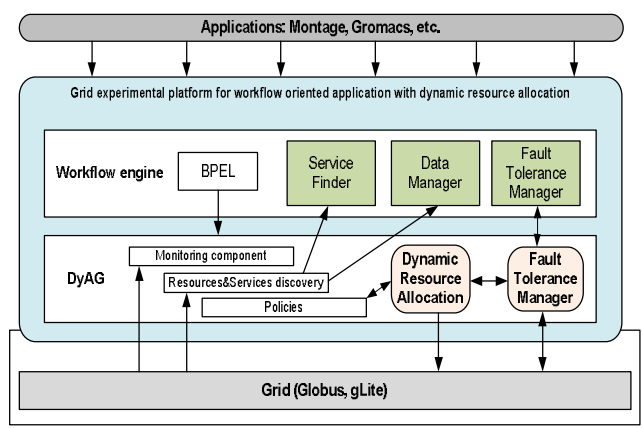

Figure 3. DyAG components

The dynamic scheduler retains a list of available resources which is kept up to date through periodic updates. This way the scheduler will always have a real time view of the Grid Environment and the available resources within it. The scheduler receives a list of invokes for various services and based on this list of resources it will search for resources where the needed services are deployed. It will then use the resources based on different implemented policies: load balancing between resources, smallest execution time or lowest data transfer.

*4.2 Fault Tolerance Manager*

Failure management is an important concern in the context of resource allocation. If a resource fails, our manager detects it in real time and the module can decide that a rescheduling is needed using the dynamic resource allocation module.

Using these mechanisms the scheduling component attempts to manage the execution of the workflow as well as possible and with minimal overhead. Due to the flexibility of the implementation, with various failures management policies which can be implemented and chosen by the user this system can handle a large number of failures which can happen during the execution of a workflow.

When referring to a fault tolerant systems (including a web services), we refer to a system which supplies a set of services to its clients, according to a well defined contract, in spite of error presence, through detecting, correcting and eliminating errors, while the systems continues to supply an acceptable set of services. A fault tolerance model highlights possible causes and conditions where errors might appear, with the goal of improving system characteristics do detect and eliminate errors. The main classes of errors that might appear in such systems are presented next (Avizienis et al., 1985).

Network errors are environmental errors caused by the communication channel and basically refer to package losses on the transmission path or corrupted incoming packages on the receiving path. These errors can be corrected by the network transmission protocol and in cases where no correction can be applied the communication path between the two endpoints is considered broken.

Timing errors are errors that can occur either at the beginning of the communication as a result of the impossibility to establish a connection, or during the communication flow when for example the response time of the called exceeds the response time expected by the caller. In case of systems which exhibit large and variable communication latencies, such timing conditions add a nondeterministic component to the expected approximate time.

Response errors are caused by a service which returns values outside of the expected boundaries by the caller. In such situations, components have to be able to validate a certain response and to appropriately handle the exceptions. A system that is designed as a state machine, can execute uncontrolled transitions in the state space which can be further propagated to other services as a result of the grid service composition.

Life cycle errors are particular to components which expose services which can expire at a certain moment. Service changes could be both syntactical and structural with different implications on the service callers.

Interaction errors are caused by incompatibilities at the communication protocol stack level, security, workflows or timing. We expect that for complex systems applications to observe a high probability of interaction error occurrence. Some of these, as for example the ones due to different security levels, could be isolated and eliminated during the testing phases in a high percentage as there is a limited number of calls between virtual organizations.

The main approach to attack fault tolerance is rollback technique (Manivannan et al, 1999), which implies application state logging at a certain time interval and restoring the last stable state in case the application is detected as entering a critical state. The used techniques are either check pointing types (Alvisi et al., 1998) where the application state is expected, or logging techniques which implies application message logging and handling.

For data grid systems, one of the most common and widespread fault tolerance techniques is provided by replication techniques, at both data provider and computing resources. In the later case, a certain application can be running in parallel on multiple resources and in case of error conditions, computation is continued on the healthy and active resources.

DyAG contains a service for fault tolerance management. It is capable to use described rollback and check pointing techniques. This service is interconnected with resource allocation module and offers information of deployed services, in the case of fault.

## 5. RELATED WORK

In this section we present a summary about interoperability between some of the workflow engines most used in scientific applications and middleware platforms. Condor DAGMan (Thain et al., 2005) submits jobs directly to the Condor scheduler; it doesn't offer support for other middleware.

Karajan provides interoperability through the use of "providers" that allow middleware selection at runtime: GT2, GT3, GT4 or Condor. It has also support for SSH protocol. Authentication is done with either user certificates (personal mode) or host certificates (shared mode).

In Taverna (Hull et al., 2006) and ActiveBPEL workflow are seen as web services. The difficulty of implementation is hidden, users are presented a high-level interface. Interoperability for Taverna is limited to MyGrid, while ActiveBpel can submit jobs to any middleware offering web services.

Triana is middleware agnostic; supports P2P, web services and Grids. GridLab GAT (Grid Application Toolkit), Triana's API for accessing Grid services, is written in such a way that new modules can be added, to achieve interoperability with different middleware platforms. Triana jobs do not have web interfaces, communication is done only through the input/output files, and submission is performed by a resource manager (GRAM1 or GRMS2). Triana can generate files entry for Pegasus / Condor.

Pegasus (Singh et al., 2005) sends its workflow to Condor DAGMAN / CondorG, in order to submit remote jobs. Pegasus users don't access DAGMan directly, except for optimization and troubleshooting.

Swift uses Globus Toolkit to submit jobs in Grid. For authentication and authorization on remote sites, it uses Grid Security Infrastructure (GSI).

P-GRADE Grid portal hides the details of low-level access to Grid resources, offering an interface which can be used with Globus Toolkit 2, Globus Toolkit 4, LCG-2 and gLite. Access to various Grids can also be done simultaneously, if the user certificates for those Grids are valid.

Many workflow engines work over a single type of middleware, besides those that enable web service orchestration (using WS-BPEL, for example) and should work with any middleware providing web services. This is one of the main reasons for choosing WS-BPEL as the specification language for our platform.

We therefore analysed from the functional point of view the existing workflow languages. We noticed that WS-BPEL and Karajan are the most complex languages supporting a large number of basic models. We chose WS-BPEL for our proposed engine since it is a standardized language that provides support for many features and it is very expressive.

## 6. CONCLUSIONS

As we have shown above, from studying the existing workflow languages and platforms we have concluded that most of the current platforms do not provide a complete coverage for aspects like abstract workflow specification, data management, fault tolerance and interoperability with multiple middleware systems.

Our goal is to develop a workflow platform that can offer all these functionalities, and we believe that the best approach for achieving this goal is to introduce an additional set of components to an existing open source workflow engine.

We chose the ActiveBPEL engine due to the fact that it is based on the WSBPEL language, which has the advantages of standardization and high expressivity, and also due to its large community of users.

The next steps in this project are to elaborate more detailed specifications for the proposed workflow engine components, to define their interface with the platform's lower and higher levels and then to start the implementation.

Performance is also an important concern, so we intend to apply a benchmark based method for comparing our platform with other similar engines.